# The Dead Internet Theory: A Survey on Artificial Interactions and the Future of Social Media

## Prathamesh Muzumdar a*, Sumanth Cheemalapati b, Srikanth Reddy RamiReddy c, Kuldeep Singh d, George Kurian e and Apoorva Muley f

*a University of Texas at Arlington, United States.*
*b Dakota State University, United States.*
*c University of Southern Mississippi, United States.*
*d Arkansas Tech University, United States.*
*e Eastern New Mexico University, United States.*
*f People's University, India.*

***Authors' contributions***

This work was carried out in collaboration among all authors. All authors read and approved the final manuscript.

**Short Communication**

## ABSTRACT

The Dead Internet Theory (DIT) suggests that much of today's internet, particularly social media, is dominated by non-human activity, AI-generated content, and corporate agendas, leading to a decline in authentic human interaction. This study explores the origins, core claims, and implications of DIT, emphasizing its relevance in the context of social media platforms. The theory emerged as a response to the perceived homogenization of online spaces, highlighting issues like the proliferation of bots, algorithmically generated content, and the prioritization of engagement metrics over genuine user interaction. AI technologies play a central role in this phenomenon, as social media platforms increasingly use algorithms and machine learning to curate content, drive engagement, and maximize advertising revenue. While these tools enhance scalability and personalization, they also prioritize virality and consumption over authentic communication, contributing to the erosion of trust, the loss of content diversity, and a dehumanized internet experience. This study redefines DIT in the context of social media, proposing that the commodification of content consumption for revenue has taken precedence over meaningful human connectivity. By focusing on engagement metrics, platforms foster a sense of artificiality and disconnection, underscoring the need for human-centric approaches to revive authentic online interaction and community building.

___

*\*Corresponding author: E-mail: prathameshmuzumdar85@gmail.com;*

## 1. INTRODUCTION

The evolution of social media platforms has significantly reshaped the way humans connect, communicate, and consume information. Emerging in the early 2000s with platforms like MySpace and Friendster, social media quickly gained traction by offering users a space to share personal updates, network with friends, and explore niche communities. With the launch of Facebook, Twitter, Instagram, and later platforms like TikTok, social media transformed from a space for personal interaction into a global phenomenon attracting billions of users. The initial success of these platforms was fueled by their ability to foster authentic connections, enable self-expression, and provide a participatory culture that encouraged active user engagement (Bakker, 2012). As social media platforms grew, their business models evolved to leverage user activity for revenue generation. The core model revolved around advertising, where platforms monetized user data to offer targeted advertisements. This approach prioritized the time users spent on the platform, as longer engagement translated directly into increased ad exposure and clicks. Platforms integrated metrics like likes, comments, and shares to measure engagement and incentivize users to remain active. As the number of subscribers and daily active users surged, so did the advertising potential, driving these platforms to focus on maximizing user interaction to optimize revenue streams.

With the exponential growth in user bases and content demands, social media platforms began incorporating artificial intelligence (AI) technologies to manage and enhance user experiences (Cook et al., 2019). AI systems curated content through algorithms, ensuring users encountered personalized feeds designed to keep them engaged. However, alongside these advancements, platforms began using AI bots to amplify interactions. Bots were programmed to like, comment, dislike, or share content, creating an illusion of heightened activity and engagement. These AI-driven interactions blurred the line between human and non-human participation, introducing a subtle yet pervasive influence on user behavior. The integration of AI bots introduced a dynamic where real users unknowingly interacted with automated systems. By stimulating discussions, mimicking trends, and generating content, bots perpetuated cycles of interaction that encouraged users to spend more time on the platform (Boeker and Urman, 2022). This increased exposure to advertisements, ultimately boosting revenue for social media companies. The intertwining of real and artificial interactions created a self-sustaining feedback loop, where the presence of bots fostered heightened activity, and this activity drove further engagement.

While this approach proved lucrative for platform owners, it also raised questions about the authenticity of online interactions and the integrity of digital spaces (Berns et al., 2021). Critics argue that the use of bots to generate engagement has contributed to a phenomenon described by the Dead Internet Theory (DIT). This theory suggests that a significant portion of online activity is no longer driven by genuine human engagement but by AI-generated content and interactions. In the context of social media, DIT highlights how platforms prioritize revenue generation over authentic social connection. By promoting content consumption and engineered engagement, social media increasingly reflects an artificial construct rather than a space for meaningful human interaction. This study aims to explore the implications of this shift, examining how AI bots and algorithmic curation have transformed social media into a machine-driven ecosystem that prioritizes profit over genuine connectivity. The findings will contextualize DIT within the evolving landscape of social media and its impact on digital culture and communication.

## 2. THEORY ORIGIN

The origins of the Dead Internet Theory (DIT) can be traced back to speculative discussions in online forums and communities in the late 2010s and early 2020s. It emerged as a response to growing unease about the changing nature of the internet, particularly in how social media and content platforms operate. Early proponents of the theory observed that much of the internet no longer felt as vibrant or genuine as it had in its earlier days, where user-generated blogs, niche forums, and personal websites created spaces for organic interaction (Cropley et al., 2010). Instead, they argued that the modern internet increasingly relied on artificial systems, including bots and AI-generated content, to sustain activity and engagement. The theory gained traction as reports surfaced about the widespread use of bots on platforms like Twitter, Facebook, and YouTube to amplify likes, comments, and shares. Coupled with the rise of sophisticated AI tools capable of generating realistic text, videos, and images, this fueled the belief that a significant portion of online activity was no longer driven by real people (DiResta & Goldstein, 2024). DIT critiques the corporatization of the internet, where platforms prioritize engagement metrics and revenue over authenticity, creating an illusion of bustling activity but masking an ecosystem where human connection is secondary to monetization strategies.

The Dead Internet Theory (DIT) suggests that the internet's organic, user-driven nature of the past has been replaced by a manufactured ecosystem dominated by artificial interactions. In its earlier days, the internet thrived on authentic content created by real users, often within niche communities and personal forums. Proponents of the theory argue that this vibrant landscape has been supplanted by an internet that is increasingly overrun by bots—automated accounts designed to mimic human interaction. These bots influence opinions, manipulate engagement metrics, and create the illusion of widespread activity (Veale and Cook, 2018). Moreover, the theory posits that a significant portion of online content today is algorithmically generated. This includes news articles, comments, reviews, and even entire websites, flooding digital spaces with AI-driven material that dilutes genuine human input. Corporations and governments are also accused of orchestrating and manipulating online spaces for financial gain, political propaganda, or social control. This alleged coordination between AI systems and powerful entities fosters a facade of dynamism and vibrancy, masking what some see as a hollow and artificial digital landscape (Walter, 2024). According to DIT, the internet may appear alive and bustling, but beneath the surface lies a construct where authentic human connection and creativity are increasingly deprioritized.

The Dead Internet Theory (DIT) is rooted in concerns about how advancements in technology and corporate practices have transformed the online landscape. A key element of the theory is the rise of AI and automation, where increasingly sophisticated AI models, such as chatbots and content generators, blur the line between human and non-human creators. Critics argue that AI now accounts for a disproportionate share of online activity, generating vast amounts of content, including articles, comments, and interactions, which mimic human behavior and create an illusion of organic activity.

The theory also highlights the decline in internet authenticity. Early platforms like forums, personal blogs, and niche websites were driven by real users engaging in unfiltered exchanges of ideas. However, the modern internet, dominated by algorithmically curated social media feeds, feels repetitive and synthetic, prioritizing content optimized for engagement rather than authenticity. Corporate overreach plays a significant role in this shift, with tech giants focusing on monetization through engagement metrics and targeted advertising. Algorithms amplify polarizing or viral content, often boosting artificial or bot-generated material over genuine user contributions. Finally, censorship and homogenization exacerbate the problem. Both governments and corporations curate online spaces to suppress dissent and promote sanitized narratives, stifling originality and creativity. These practices reinforce the perception that the internet, while seemingly active, lacks genuine human connection and diversity.

## 3. SOCIAL MEDIA AND CONTENT FARMING

Social media platforms have increasingly adopted content farming strategies to maximize user engagement and capitalize on network effects through what is often termed an "architecture of participation." This model encourages continuous user interaction by designing systems that promote content creation and sharing, thus amplifying engagement metrics. Central to this strategy is the shift from chronological content presentation to algorithmically curated feeds. By prioritizing content that aligns with a user's perceived interests, recommendation algorithms create a personalized experience designed to keep users scrolling and interacting. However, this curation often emphasizes popular, sensational, or emotionally charged content, overshadowing diverse or less engaging material.

Content farming, driven by these algorithms, produces large volumes of repetitive or low-quality material aimed at gaming the system. This content is optimized for visibility within the algorithmic framework, often prioritizing quantity over quality to capture user attention. Consequently, while these practices enhance platform engagement and ad revenue, they also lead to the homogenization of online spaces, suppressing unique voices and authentic interactions. The result is an ecosystem where content is not necessarily served based on relevance or originality but on its ability to drive interaction, reinforcing cycles of engagement that prioritize profit over meaningful user experiences.

## 4. AI AND CONTENT FARMING

The use of AI for content farming has transformed how digital platforms generate and manage online material. AI-powered tools such as language models, image generators, and video production systems enable the rapid creation of vast quantities of content tailored to engage users and optimize platform metrics. These technologies can produce clickbait articles, auto-generated blogs, and even social media posts that align with trending topics or specific user interests, effectively fueling engagement while minimizing human effort. The scalability of AI allows platforms to flood feeds with relevant material, ensuring users remain immersed and exposed to advertisements, which are critical to revenue generation.

While AI-driven content farming increases efficiency and engagement, it also raises concerns about authenticity and quality. Automated systems often prioritize volume over substance, resulting in a proliferation of repetitive or superficial material that saturates digital spaces. Furthermore, these algorithms are programmed to exploit user psychology by generating emotionally charged or sensational content to maximize clicks and shares. This approach not only erodes trust in online platforms but also suppresses original, human-created material, as AI-generated content dominates visibility in algorithmically curated feeds. The widespread use of AI in content farming reflects a shift toward prioritizing engagement metrics over meaningful, high-quality interactions and contributions online.

## 5. THEORETICAL EVIDENCE IN SOCIAL MEDIA

Several pieces of evidence lend support to the Dead Internet Theory (DIT), suggesting that much of today's online activity is driven by non-human forces. Bot activity plays a significant role, with studies estimating that 40% to 60% of web traffic is generated by bots. These automated systems engage in scraping, spam, and manipulation, creating artificial interactions that mimic genuine human activity. Bots are also frequently employed to inflate metrics, such as likes, shares, and comments, fostering the illusion of vibrant online engagement.

The AI content boom further supports the theory. Platforms like Reddit, YouTube, and Twitter/X have experienced a surge in repetitive, low-quality, or spam-like material, often attributed to AI-driven generators. This influx of algorithmically created content reduces the space for authentic human contributions, leaving the internet cluttered with indistinct and superficial material. Additionally, the proliferation of echo chambers exacerbates this issue. The repetitive nature of viral trends, memes, and news across platforms suggests a lack of diversity and creativity, reinforcing perceptions that online content is increasingly homogenized. Astroturfing—or the creation of fake grassroots campaigns by corporations or governments—further undermines the authenticity of digital spaces. These manipulative efforts influence public opinion and create the illusion of organic support, amplifying the idea that much of the internet's activity is artificial and controlled.

## 6. DEFINING DEAD INTERNET THEORY

From the perspective of social media, the Dead Internet Theory (DIT) can be redefined as the idea that modern online platforms have transitioned from spaces of genuine human interaction to ecosystems dominated by artificial activity, primarily driven by bots, AI-generated content, and corporate algorithms. Social media, once heralded as a medium for authentic connection and community building, now prioritizes engagement metrics, ad revenue, and algorithmic optimization over fostering meaningful human relationships.

In this reimagined digital landscape, a significant proportion of activity—such as likes, comments, shares, and even content creation—is orchestrated by bots and AI systems designed to mimic human behavior. These artificial agents inflate metrics, manipulate trends, and generate repetitive, low-quality content that dilutes authentic user participation. The overwhelming presence of AI-driven interactions creates a feedback loop, keeping real users engaged through curated content but at the expense of genuine social connectivity. Moreover, social media platforms have evolved into tightly controlled environments where corporations and, in some cases, governments exert influence through algorithmic curation and censorship. This results in sanitized, homogenized content that suppresses dissent and originality. Under this framework, the "dead" internet is not devoid of activity but is lifeless in its human essence, prioritizing consumption over connection and profit over authenticity.

## 7. CONSEQUENCES OF THE THEORY

The consequences of the Dead Internet Theory (DIT) are multifaceted, presenting both positive and negative implications. On one hand, the theory highlights concern about the decline of authenticity in online spaces, where interactions and content are increasingly driven by algorithms and bots rather than genuine human effort. This shift may lead to reduced trust in social media platforms, as users grow wary of artificial engagement and question the authenticity of trends, reviews, and discussions. Additionally, the homogenization of content, driven by algorithmic curation, stifles creativity and diversity, making online spaces feel repetitive and uninspired.

However, some consequences of DIT reveal potential benefits and alternative perspectives. While bot traffic is significant, genuine human interaction continues to thrive, with billions of users actively engaging in forums, social media, and collaborative projects. This resilience of human activity underscores the internet's role as a global connector. Critics also argue that DIT romanticizes the early internet, overlooking its challenges, including scams, spam, and bad actors, which were present even in its so-called "golden age." Furthermore, the rise of automation and AI could reflect technological advancements enabling mass participation and efficient content creation, democratizing access to information and tools that were once limited to select users. This duality underscores the complexity of DIT's implications.

## 8. DEAD INTERNET THEORY-SOFTWARE ARTIFACT

Fig. 1 illustrates the proposed artifact based on current literature regarding the use of AI for content creation and disposition. The workflow diagram is organized into four levels: the User Level at the front end, and three backend levels — the Query Aggregator Level, the Content Creation Level, and the Profile Cluster Level. The workflow follows a top-down pattern. It begins when users (profiles) create a primary query in the form of textual posts. These queries are then sorted and grouped into buckets at the Query Aggregator Level.

This sorting process leverages text analytics models and a sorting algorithm to segregate keywords effectively. At the Content Creation Level, the grouped queries are processed by a Generative AI-based autonomous text creator to generate content related to the primary query. This stage also includes manual intervention by human moderators to ensure content quality. The Content Creation Level is particularly intricate, comprising multiple text generation and validation models integrated into a recursive and backtracking algorithm.

The Profile Cluster Level clusters users submitting similar queries into pools using a keyword-based text analytics segregation model. The model can be further refined with additional parameters such as demographics or geographic data. Finally, post bots deliver the generated content to the appropriate segregated profiles. This workflow represents a software artifact built upon diverse algorithmic structures, including recursive algorithms, backtracking, sorting, divide-and-conquer, and dynamic programming techniques.

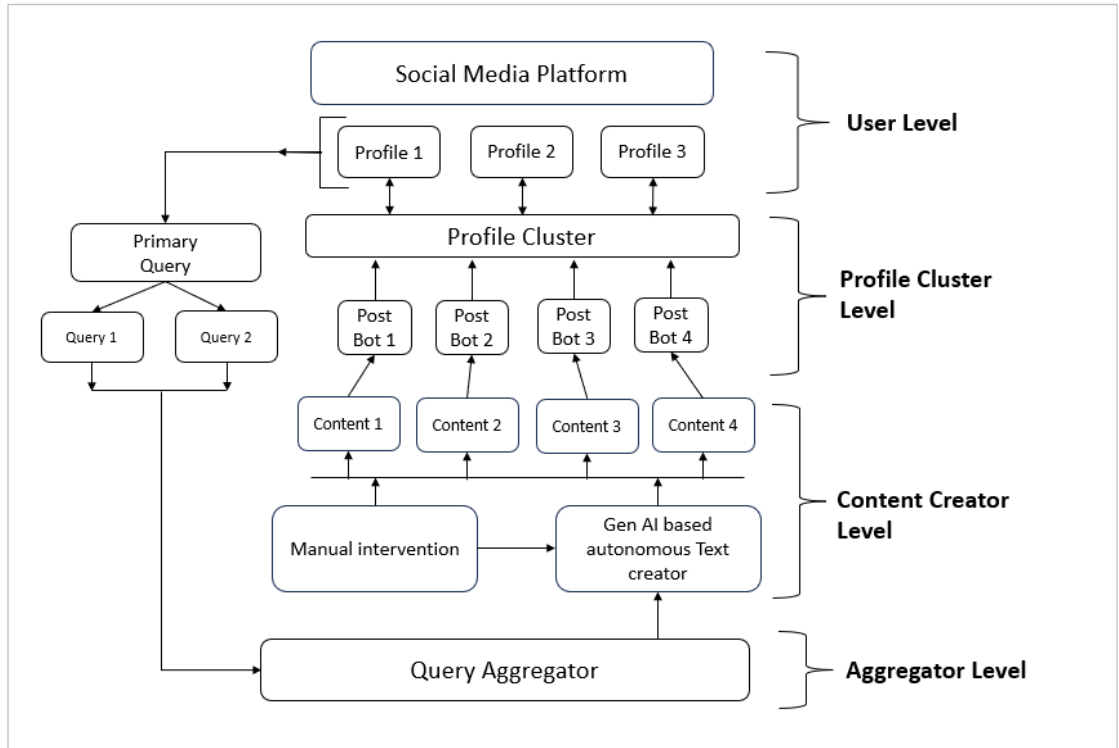

**Fig. 1. Software Artifact- AI automated bot workflow diagram**

## 9. CONCLUSION

In conclusion, the Dead Internet Theory (DIT) underscores significant transformations in the digital landscape, particularly within social media. While the theory raises valid concerns about the growing presence of AI-driven content, bots, and corporate influence diluting online authenticity, it also reflects nostalgia for a less commercialized internet. Despite the challenges of homogenization and artificial engagement, human activity remains a driving force in digital spaces, bolstered by evolving technology enabling mass participation. DIT invites critical reflection on the balance between technological innovation and preserving genuine human connection, emphasizing the need for platforms to prioritize authenticity alongside profitability.

## DISCLAIMER (ARTIFICIAL INTELLIGENCE)

Author(s) hereby declares that NO generative AI technologies such as Large Language Models (ChatGPT, COPILOT, etc.) and text-to-image generators have been used during the writing or editing of this manuscript.

## COMPETING INTERESTS

Authors have declared that no competing interests exist.

*Peer-review history:*
*The peer review history for this paper can be accessed here:*
*https://www.sdiarticle5.com/review-history/128383*